\documentclass[%
superscriptaddress,
 nofootinbib,
 amsmath,amssymb,
 aps,
 notitlepage,
 prl,
 twocolumn,
 noeprint
 nobalancelastpage,
 longbibliography,
]{revtex4-2}

\usepackage{graphicx}%
\graphicspath{{Figures/}}
\usepackage{dcolumn}%
\usepackage{bm}%
\usepackage[usenames,svgnames]{xcolor}
\usepackage[
    colorlinks,
    citecolor=DarkBlue,
    linkcolor=Maroon,
    urlcolor=DarkGreen,
    ]{hyperref}%

\usepackage{microtype}
\usepackage[mathcal]{euscript}
\usepackage{comment}
\usepackage{subcaption}
\usepackage[capitalise,nameinlink]{cleveref}

\usepackage[
]{siunitx}

\newcommand{\rmd}{\mathrm{d}}
\newcommand{\rme}{\mathrm{e}}
\newcommand{\rmi}{\mathrm{i}}
\newcommand{\tx}[1]{\text{(#1)}}

\DeclareMathOperator{\real}{Re}
\DeclareMathOperator{\imag}{Im}
\DeclareMathOperator{\erf}{erf}

\hypersetup{
    pdftitle = {Searching for dark matter via modulation of fundamental constants using the reference cavity in Advanced LIGO}
  }

\begin{document}

\preprint{APS/123-QED}

\title{Advanced LIGO, LISA, and Cosmic Explorer as dark matter transducers}

\author{Evan Hall}
\affiliation{LIGO Laboratory, MIT Kavli Institute, Department of Physics, MIT, Cambridge, MA 02139, USA }
\author{Nancy Aggarwal}
\affiliation{Center for Fundamental Physics, Department of Physics and Astronomy,}
\affiliation{Center for Interdisciplinary Exploration and Research in Astrophysics (CIERA), Northwestern University, Evanston, IL 60208, USA}

\newcommand{\NA}[1]{{\color[RGB]{200,0,250}#1}}
\newcommand{\edh}[1]{{\color[RGB]{0,30,150}#1}}

\newcommand{\visiblecomment}[1]{{\color[RGB]{200,0,0}#1}}
\date{\today}

\begin{abstract}
    {We present a method to search for scalar field ultralight dark matter directly interacting with gravitational-wave interferometers via a modulation of the fine structure constant and the electron mass. This modulation induces an effective strain in solid materials at a frequency determined by the mass of the dark matter particle. We study the prospects for looking for such an effect in the LIGO detectors by using the solid cavity which is nominally used for pre-stabilizing the laser frequency and we project upper limits. We contextualize them with previous limits from GEO600, possible limits from a similar strain in the LIGO beamsplitter, and with potential limits from upcoming experiments like LISA, Cosmic Explorer and from an upgraded solid cavity. We find that with the sensitivity of Advanced LIGO, competitive upper limits on DM coupling can be placed at the level of $\left\vert d_{m_e}+d_e\right\vert \sim 0.2$ for $m_\text{DM} \sim 10^{-13}\,\mathrm{eV}/\mathrm{c}^2$ with a combination of two searches using the solid cavity and the beamsplitter in LIGO; future experiments could reduce this upper limit to $\sim10^{-3}$.}
\end{abstract}

\maketitle

\section{Introduction}

Dark matter (DM) is known to interact gravitationally with normal matter, but so far, numerous efforts to look for a secondary interaction have not come to fruition~\cite{rubin1983dark,Planck:2018vyg,Clowe:2006eq,Bertone:2004pz}.
The attometer precision of modern Michelson interferometers makes them suitable to look for gravitational waves (GWs), but they can also be used to search for various dark matter candidates, either via direct interaction with the interferometer~\cite{Hall:2016usm,Pierce:2018xmy,Nagano:2019rbw,Guo:2019ker,Michimura:2020vxn,Nagano:2021kwx,Baum:2022duc,LIGOScientificCollaborationVirgoCollaboration:2021eyz}, or by generation of GWs by axion DM~\cite{Arvanitaki:2014wva,Baryakhtar:2017ngi,Arvanitaki:2016qwi,Sun:2019mqb,Yuan:2022bem} or primordial black holes~\cite{Sasaki:2016jop,LIGOScientific:2021job,Nitz:2021vqh,Miller:2021knj,Jedamzik:2020omx}. Furthermore, mapping out black-hole populations and merger rates will also provide clues regarding the nature of DM~\cite{Arvanitaki:2014wva,Baryakhtar:2017ngi,Ng:2020ruv,Jedamzik:2020omx,Clesse:2016vqa,Gow:2019pok,Hutsi:2020sol,Franciolini:2021tla}. %

Ultralight scalar fields -- another promising class of candidate for DM --  can also cause a direct interaction with GW detectors~\cite{Arvanitaki:2014faa,Stadnik:2014tta,Grote:2019uvn,Morisaki2019detectability}.
When coupled to the standard model, such a field modulates the fine-structure constant \(\alpha\) and the mass of the electron \(m_e\), in turn leading to a periodic strain in solid materials due to a change in atomic radius ~\cite{Arvanitaki:2014faa,Stadnik:2015kia,Arvanitaki:2015iga}. 
For a monochromatic DM signal, the modulation of the fine structure constant \(\alpha\) and the mass of the electron \(m_e\) ~\cite{Arvanitaki:2014faa,Stadnik:2014tta} 
\begin{align}
    \frac{\Delta\alpha(t)}{\alpha} &= A_e \cos\left(\Omega_\text{DM} t\right)\\
    \frac{\Delta m_{e}(t)}{m_e} &= A_{m_e} \cos\left(\Omega_\text{DM} t\right)
\end{align}
occurs at an angular frequency
\(
    \Omega_\text{DM} = m_\text{DM}c^2/\hbar,
\)
where \(m_\text{DM}\) is the mass of the DM particle.
The strength of the modulation is given by $A_{e,m_e} \sim d_{e,m_e}\frac{\hbar}{m_\text{DM}c^3}\sqrt{8\pi \varrho_\text{DM} G}$,
where \(\varrho_\text{DM}\) is the local DM energy density and \(d_{e}\) and \(d_{m_e}\) are the DM couplings with Standard Model~\cite{Geraci:2018fax}.
This implies a typical strain $A \sim \num{7e-18} \times d$ for $m_\text{DM} = 0.1\, \mathrm{peV}/\mathrm{c}^2$.
Under the cold dark matter scenario, the DM is not exactly monochromatic but is expected to have a narrow linewidth:  \(\Delta \Omega_\text{DM} / \Omega_\text{DM} \sim  10^{-6}\)~\cite{Derevianko:2016vpm}. 

This modulation of fundamental constants will lead to a strain in a solid body, which in the non-relativistic limit can be simplified to~\cite{Arvanitaki:2015iga}
\begin{equation}
   {h}_\text{DM}(t) = -\left(\frac{\Delta\alpha(t)}{\alpha} + \frac{\Delta m_{e}(t)}{m_e} \right).
\end{equation}
This strain can be measured by a resonant mass, as was done in the AURIGA GW detector~\cite{Branca:2016rez}. Separately, such a DM strain can also appear in a free-space Michelson interferometer if its two arms contain unequal amounts of solid material, which can happen when light in one arm traverses the substrate material of the beamsplitter optic~\cite{Grote:2019uvn}; this effect was recently exploited in the GEO600 and Holometer interferometers to search for scalar dark matter~\cite{Vermeulen:2021epa,Aiello:2021wlp}.
This effect can also be detected by locking a laser to an optical cavity made out of a solid material and monitoring the frequency modulation of the laser by, for example, comparing it to a laser locked to a free-space suspended cavity~\cite{Geraci:2018fax,Derevianko:2016vpm}.
In this scheme, the suspended cavity acts as the stable frequency reference, and the solid cavity is the primary DM signal transducer. We note that these same DM fields can also couple to the masses of objects (in addition to their size) via a direct coupling to the gluons, the effect of this coupling with GW interferometers was studied in Refs.~\cite{Arvanitaki:2014faa,Morisaki2019detectability}.

\begin{figure}
    \centering
    \includegraphics[width=\columnwidth]{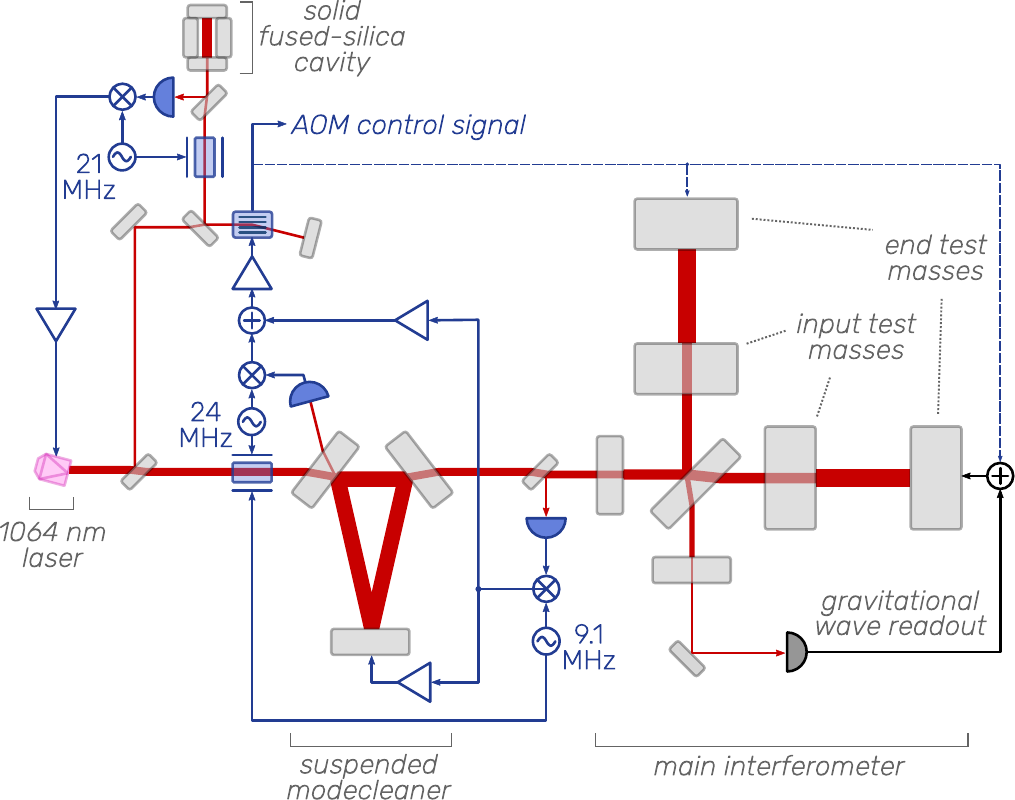}
    \caption{Advanced LIGO frequency stabilization.
    Noise in the solid cavity, including noise from any dark matter signal that changes the length of the solid cavity, will appear on the control signal applied to the acousto-optic modulator (AOM).
    The various noise contributions to this control signal are given in \cref{eq:aom fluctuations}.
    \label{fig:carm_diagram}}
\end{figure}

We propose to constrain the coupling strength of above DM models by inspecting the relative frequency changes between a solid cavity and two suspended cavities in Advanced LIGO. We also investigate the prospects of monitoring the beamsplitter thickness in Advanced LIGO.
Furthermore, we propose cross correlating the two LIGO detectors for a stronger suppression of spurious environmental effects.

In the second section of this paper, we describe how the reference cavities~\cite{Chalermsongsak:2014aua} in the current LIGO detectors can be used as DM monitors.\footnote{When using the phrase \textit{reference} cavity, we will mean the solid cavity.}
In the third section, we compare this method to other ground-based schemes. 
We show that at lower DM masses, the Advanced LIGO reference cavity method can perform better than the GEO600 beamsplitter method. At intermediate masses, we find that the Advanced LIGO beamsplitter constraints will outperform GEO600 and the Advanced LIGO reference cavity  constraints.
In the fourth section, we discuss the prospects of using the reference cavity in LISA in a similar manner to constrain lower-mass scalar fields.

\section{LIGO reference cavity}

Advanced LIGO's frequency stabilization is shown in \cref{fig:carm_diagram}, showing a system of nested frequency locking loops involving the fused silica reference cavity, a suspended triangular modecleaner, and the main interferometer, which is also formed from suspended optics~\cite{2012OExpr..2010617K,LIGOScientific:2014pky,Bode:2020dge}.
In brief, the laser frequency is first stabilized to the reference cavity using Pound--Drever--Hall (PDH) locking, with a typical loop bandwidth in excess of \SI{100}{kHz}.
It is then additionally stabilized to the suspended modecleaner, again with PDH locking and with a bandwidth of about \SI{50}{\kHz}, by feeding the PDH control signal to a double-pass acousto-optic modulator (AOM) placed at the input of the reference cavity.
Finally, the laser light is PDH stabilized to the common-mode arm degree of freedom of the main interferometer with a bandwidth of tens of kilohertz by feeding the PDH control signal to the error point of modecleaner's frequency locking loop~\cite{Cahillane:2021jvt}.
At timescales slower than a few hertz, the AOM control signal is fed back to the main interferometer test masses to prevent the accumulation of large seismic- and tidally-driven offsets (dashed line).
The timescale of this tidal servo sets a lower frequency limit for using the AOM control signal to search for DM-induced fluctuations in the reference cavity, while the bandwidth of the suspended modecleaner sets an upper frequency limit to the search.

Overall, the effect of the nested frequency locking system is that any relative frequency noise between the reference cavity and the suspended optics appears in the control signal $a(t)$ applied to the double-pass AOM at the reference cavity input.
In particular, a dark matter signal that induces a solid-body strain $\mathfrak{h}_\text{DM}$ will produce a frequency modulation in the reference cavity $\Delta\nu / \nu = \Delta L_\text{rc} / L_\text{rc} = \mathfrak{h}_\text{DM}$.
The same signal will also produce strains in the various suspended optics, but this signal is suppressed by $\ell / L$, where $L$ is the length of the suspended cavity and $\ell$ is the effective thickness of the suspended mirrors.
For the modecleaner $\ell / L$ is of order $10^{-3}$, and for the main interferometer it is $10^{-4}$.
Therefore, any dark matter strain in the solid cavity will appear in the feedback control signal applied to the double-pass AOM.\footnote{One could also look for a strain in the feedback control from the main interferometer to the suspended modecleaner, which typically has lower frequency noise than the solid cavity control signal, but such a search is not as sensitive due to the $\mathcal{O}(10^{-3})$ signal suppression mentioned above.}

\subsection{System noises}

In the frequency domain, the feedback control applied to the double-pass AOM is approximately
\begin{multline}
\label{eq:aom fluctuations}
    a(\Omega) \simeq -\nu_0 \times {h}_\text{DM}(\Omega) - n_\text{rc}(\Omega) - \frac{n_\text{laser}(\Omega)}{G_\text{rc}(\Omega)} \\ - \frac{n_\text{mc}(\Omega)}{G_\text{ifo}(\Omega)} + n_\text{ifo}(\Omega);
\end{multline}
here %
all noise terms $n$ are referred to frequency fluctuation; $\nu_0 = c/\lambda_0$ is the laser frequency, with $\lambda_0 = \SI{1064}{\nm}$ being the laser wavelength.
The quantity $n_\text{rc}$ describes noises associated with the (apparent) length of the solid cavity, other than the noise from a DM signal.
These other noises include thermal fluctuations and shot noise.
The quantity $n_\text{laser}$ is the intrinsic (free-running) frequency fluctuation of the laser, which is suppressed by the loop gain $G_\text{rc}$ of the solid cavity laser locking loop.
The quantity $n_\text{mc}$ describes noises associated with the (apparent) length fluctuation of the suspended modecleaner, similar to the solid reference cavity, and $n_\text{ifo}$ similarly describes noise associated with the main interferometer.
Assuming each of these contributions to $a$ is uncorrelated with the others, the resulting power spectral density of the feedback control signal is thus
\begin{multline}
\label{eq:aom psd}
    S_{a}(\Omega) = \nu_0^2 S_\text{DM}(\Omega) + S_\text{rc}(\Omega) + S_\text{laser}(\Omega)/|G_\text{rc}(\Omega)|^2 \\ + S_\text{mc}(\Omega)/|G_\text{ifo}(\Omega)|^2 + S_\text{ifo}(\Omega).
\end{multline}

\begin{table}[t]
\caption{\label{tab:refcav params}Parameters deciding thermal noise in solid reference cavities.
The current reference cavity parameters are as-built~\cite{Chalermsongsak:2014aua}. ``Upgraded'' parameters are proposed with a modest, achievable upgrade.}
\begin{ruledtabular}
    \begin{tabular}{r c l l}
    Cavity parameter & Symbol   & Current  & Upgraded \\
    \hline
    Length [cm] & $L_\text{rc}$ & 20.3 & 30.0 \\
    Beam size [mm]& $w$ & 0.29 & 3.0 \\
    Coating loss & \(\phi\) & \num{4.4e-4}  & \num{2.2e-4} \\ 
    Substrate/spacer loss & \(\phi_\text{s}\) & \num{1e-7} & \num{1e-7} \\
    Coating thickness [\textmu{}m] & \(d\)  & 4.5 & 4.5 \\
    Wavelength [nm] & \(\lambda_0\) & 1064 & 1064 \\
    Temperature [K] & \(T\)& 300 & 300 \\
    Young modulus [GPa] &\(E\) & 72 & 72\\
    Poisson ratio &\(\sigma\) & 0.17 & 0.17 \\
    Input power [mW] & \(P_\text{rc}\) & 10 & 10 \\
    Finesse & \(\mathcal{F}_\text{rc}\) & \(10^4\) & \(10^4\)
    \end{tabular}
\end{ruledtabular}
\end{table}

Many of the contributions to $S_\text{rc}$ for the LIGO reference cavities have been measured in a laboratory setting~\cite{Chalermsongsak:2014aua}.
An irreducible contribution to $S_\text{rc}$ is the thermal noise in the cavity, particularly the Brownian motion of the mirror coatings.
As can be determined from parameters in \cref{tab:refcav params} and the expression in supplemental material (SM), two mirrors of the cavity should together yield a Brownian noise limit of 
\(
 \left(\SI{4}{\mHz\big/\sqrt{\Hz}}\right) \times \sqrt{{2\pi{\times}\SI{100}{\Hz}}\,/\,{\Omega}}
\). %
Additional Brownian noise contributions come from the mirror substrates and from the cavity spacer, and at low frequencies the noise from thermoelastic damping of the mirror substrates becomes comparable to the Brownian noise level.
Aside from thermal noise, the typical shot noise limit is $\SI{e-4}{\Hz\big/\sqrt{\Hz}}$ (ref SM and \cref{tab:refcav params}), which is below the thermal noise level.
Additionally, \citet{Chalermsongsak:2014aua} measured $S_\text{laser}/|G_\text{rc}|^2$ and found it negligible compared to the thermal noise level below \SI{5}{\kHz}.
Also there is the noise of the voltage-controlled oscillator driving the AOM, which has a broadband noise level of about \SI{e-4}{\Hz\big/\sqrt{\Hz}}~\cite{Angert2009}.
Finally, the digital readback of $a(t)$ has an associated noise floor, although this digitization does not affect the performance of the servo loop.

%
\begin{figure}
    \centering
    \includegraphics[width=\columnwidth]{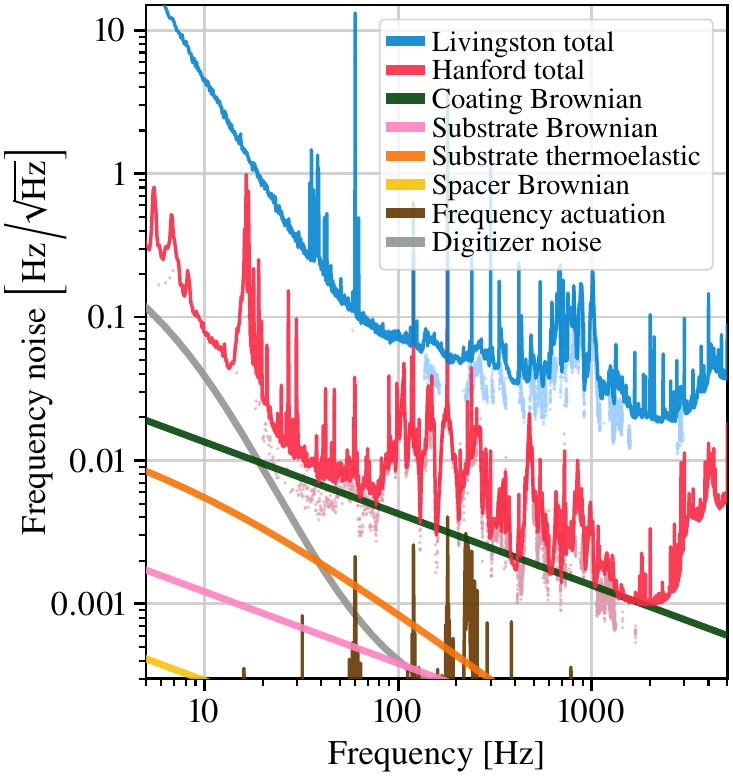}
    \caption{Feedback control noise $\sqrt{S_a(\Omega)}$ to the solid cavities at Hanford and Livingston (\cref{eq:aom psd}), along with the expected noises from Brownian motion of the cavity mirror coatings, frequency actuator noise, and digitizer noise.
    Contributions from laser table vibration are shown as faded dots underneath the measured curves.
    }
    \label{fig:IMC-F}
\end{figure}

\begin{figure}
    \centering
    \includegraphics[width=\columnwidth]{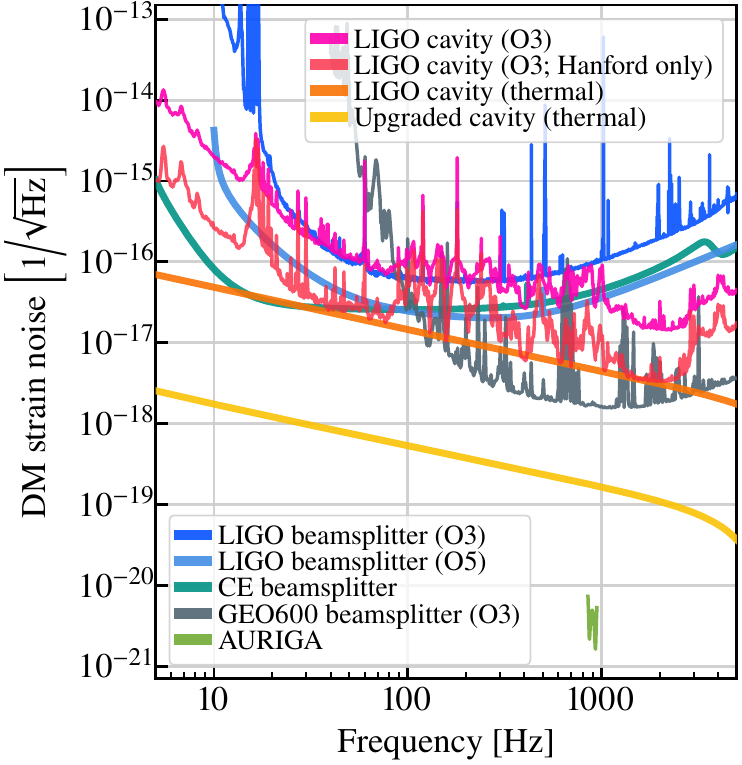}
    \caption{
    Solid-body strain noise in ground-based gravitational-wave experiments. The bright purple trace shows the limit from cross-correlating the frequency-stabilizing solid cavities of LIGO Hanford and LIGO Livingston.
    The rose-pink trace shows the same limit if the LIGO cavities were limited by thermal noise only.
    The pale orange trace shows the thermal noise limit for a low-noise and longer spacer cavity.
    The cool color traces are limits from strains in the beamsplitters of the main interferometers, and the sensitivity of the resonant bar AURIGA is also shown.
    The binwidths for these spectral densities are much higher than the binwidths that would be used for a DM search, so the value of unresolved peaks in these spectra do not faithfully represent the search sensitivity at those frequencies.
    }
    \label{fig:DM strain spectra}
\end{figure}

In \cref{fig:IMC-F}, we show the measured noise in the reference cavity loops at Livingston and Hanford during the third observing run (O3), as well as the expected thermal noise limit.
We also show the contribution from digitizer noise, laser table vibrations, and frequency actuation noise.
In \cref{fig:DM strain spectra} we show the DM-strain-equivalent of the solid cavity noise, as a geometric mean of the Livingston and Hanford noises.%

\subsection{Projected limits}

For the LIGO Hanford and Livingston detectors, the signal ${h}_\text{DM}$ is common to both since the coherence length of the dark matter is $\lambda_\text{DM} = \hbar / m_\text{DM} v_\text{DM} = c^2 / \Omega_\text{DM} v_\text{DM}$, which is at least \SI{10000}{\km} for $\Omega_\text{DM} / 2\pi < \SI{5}{\kHz}$.
Such a signal could be uncovered via a cross-correlation search assuming that the shape of the dark matter strain spectral density $S_\text{DM}(\Omega)$ is known.
The optimal signal-to-noise ratio $\rho$ for such a search is~\cite{Romano:2016dpx,Foster:2020fln}
\begin{equation}
\label{eq:rhosq_stochastic}
\rho^2 = 2T \int\limits_0^\infty \frac{\mathrm{d}\Omega}{2\pi} \frac{S_\text{DM}(\Omega)^2}{S_\text{H}(\Omega) S_\text{L}(\Omega)}
\end{equation}
with $S_\text{H}(\Omega)$ being the power spectral density of the Hanford strain-referred frequency control feedback signal $a_\text{H}(t) / \nu_0$, and similarly for Livingston (\cref{eq:aom fluctuations,eq:aom psd}).
$T$ is the total time of the search.

The dark matter strain spectral density is given by $S_\text{DM}(\Omega) = 4 \pi \mathfrak{h}_\mathrm{DM}^2 F(\Omega)$, where \(\mathfrak{h}_\mathrm{DM} = \bigl\lvert A_e + A_{m_e} \bigr\rvert  \) and $F(\Omega)$ is the dark matter lineshape.
This lineshape has been computed by, e.g., \citet{Derevianko:2016vpm}, under the standard assumptions that the dark matter halo is a coherently oscillating field with thermal distribution of modes (see SM).
The lineshape is narrowly peaked near $\Omega = \Omega_\text{DM}$, meaning that the spectra $S_\text{H}(\Omega)$ and $S_\text{L}(\Omega)$ in \cref{fig:IMC-F} can be approximated by their values at $\Omega_\text{DM}$.
Then using the result $\int_0^\infty (\rmd\Omega /2\pi) \, S_\text{DM}(\Omega)^2 = \sqrt{\pi} [\erf(\eta)/\eta] \, \mathfrak{h}_\mathrm{DM}^{\,4} \tau_\text{DM}$ (see SM),
the achievable limit on the detectable dark matter strain is
\begin{equation}
    \label{eq:dm_limit_xcorr}
    \mathfrak{h}_\mathrm{DM}   \le \left(\frac{\eta}{\sqrt{\pi}\erf{\eta}} \, \frac{\rho^2 S_\text{H}(\Omega_\text{DM}) S_\text{L}(\Omega_\text{DM})}{2 T \tau_\text{DM}}\right)^\frac{1}{4}\!\!\!,
\end{equation}
where, \(\eta = v_\text{gal} / v_\text{DM}\) is the ratio of the velocity of the solar system and the Maxwell-Boltzmann velocity of the DM field, and \(\tau_\text{DM} = c^2\big/v_\text{DM}^2\Omega_\text{DM}\) is the coherence time of the DM signal.
We take $\eta \sim 1$ and \(v_\text{DM}/c \sim 10^{-3}\).
\cref{eq:dm_limit_xcorr} is valid so long as $T > \tau_\text{DM}$.
\begin{figure*}
    \centering
    \includegraphics[width=0.357\linewidth]{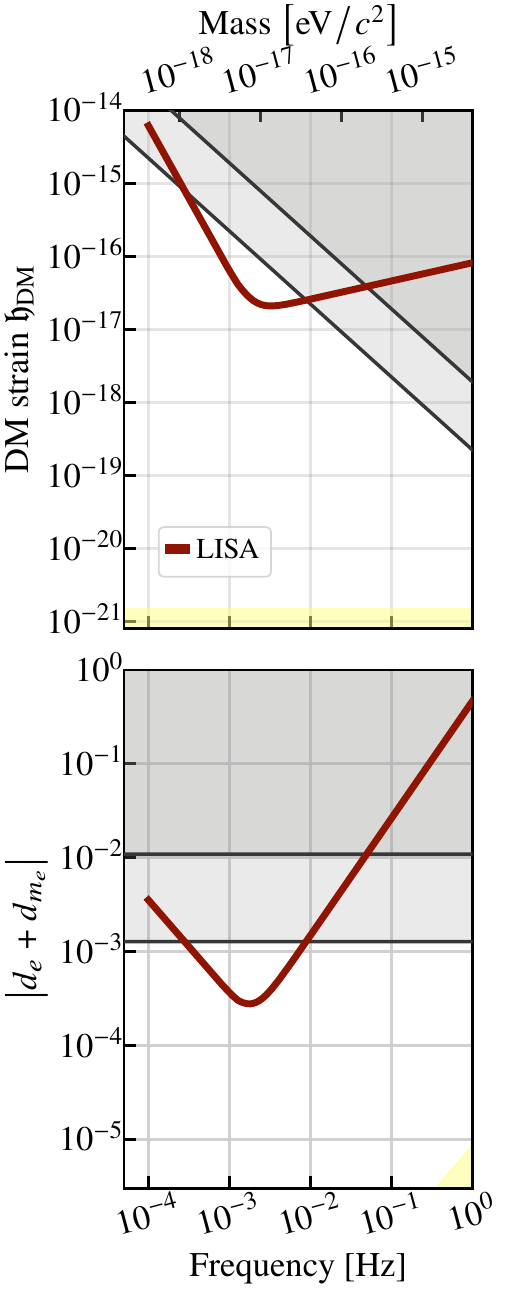}\includegraphics[width=0.643\linewidth]{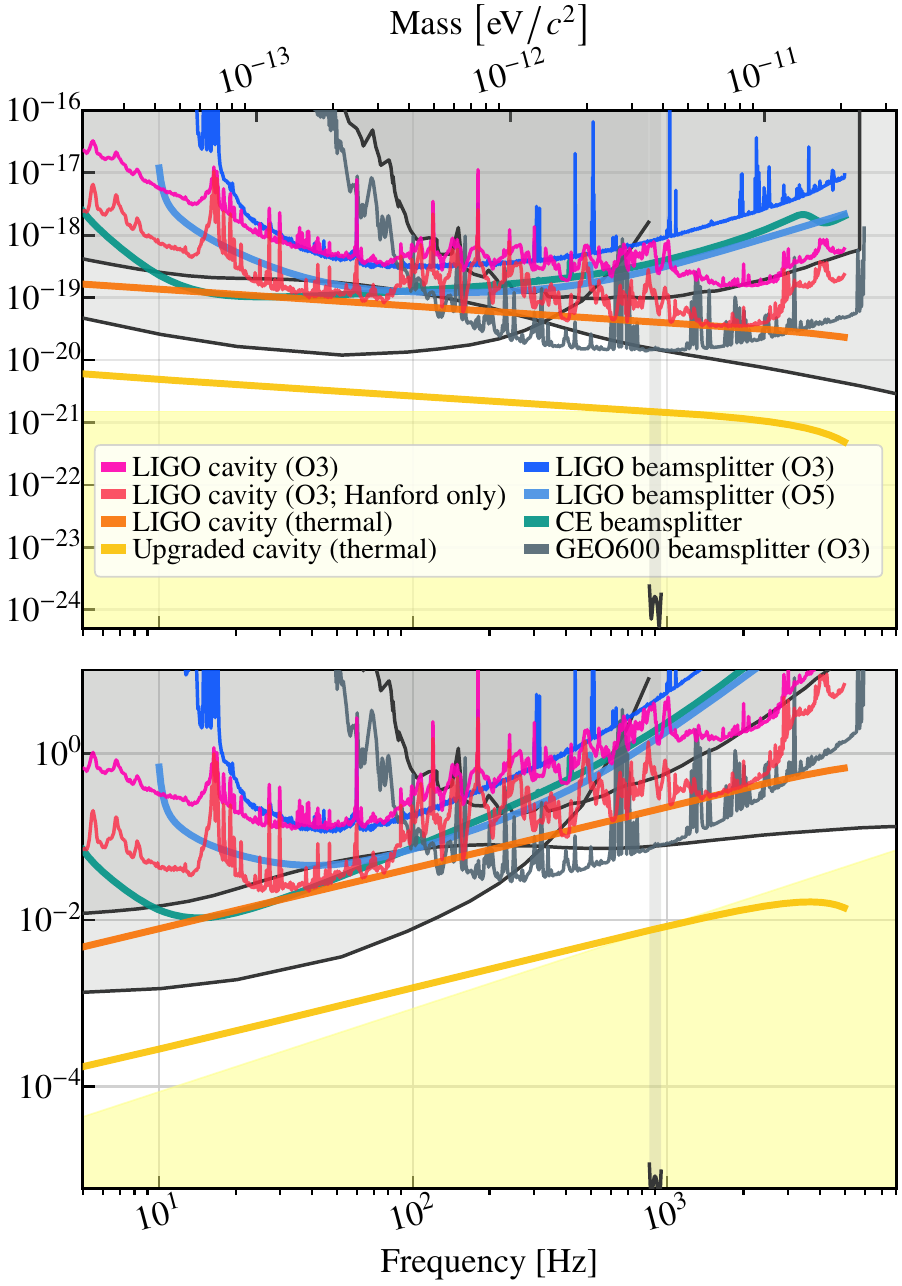}
    \caption{Traces from \cref{fig:DM strain spectra} recast as DM strain limits (Top) and limits on $\left\lvert d_e + d_{m_e} \right\rvert$ (Bottom). Left: LISA  assuming an integration time  $T = \SI{4}{yr}$; Right: ground based experiments with  $T  = \SI{1000}{\hour}$.
    In all cases a threshold signal-to-noise ratio \(\rho = 3\) is assumed. The gray regions with black lines show the existing limits from tests of the equivalence principle using torsion balances~\cite{Adelberger:2003zx} and satellite missions~\cite{Berge:2017ovy},\protect\footnote{We follow Refs.~\cite{Arvanitaki:2014faa,Arvanitaki:2015iga}, which set \(d_{m_e}\gg\left\{d_e,d_g,d_{m_u},d_{m_d}\right\}\). If reinterpreted as a constraint on $d_e$ only, the constraint is about one order of magnitude stronger.} from \SI{28}{\hour} of GEO600 data~\cite{Vermeulen:2021epa}, and from AURIGA~\cite{Branca:2016rez,AxionLimits}.
    The lower yellow space shows the region of natural values of \(d_{m_e}\) corresponding to a \SI{10}{\tera\eV} scale.}
    \label{fig:DM strain}
\end{figure*}

In \cref{fig:DM strain} we show the projected upper limit from LIGO's solid cavities%
, as well as the thermal noise limit, in both strain as well as limits on the coupling constant \(d\).
We assume an integration time of 1000 hours and signal-to-noise ratio threshold \(\rho\) of 3~\cite{Romano:2016dpx}.
Because of the look-elsewhere effect, the exact threshold required to achieve a specified statistical significance depends on the number of frequency bins that are searched over~\cite{PalkenThesis}.

\section{Other ground based experiments}

We now summarize other possible ways of searching for this effect in ground-based experiments. Firstly, we study the effect of upgrading the LIGO reference cavity to a longer spacer and lower loss amorphous coatings as shown in \cref{tab:refcav params}. We show the spacer and coating thermal noise limit in \cref{fig:DM strain spectra} %
and the potential upper limits from this upgrade in \cref{fig:DM strain}.%

Another way to constrain this DM model is by looking for solid-body strain experienced by the beamsplitter of a Michelson interferometer. %
Given the GW strain noise $h_\text{GW}(t)$ in a Michelson interferometer, its arm length $L_\text{arm}$, and the effective thickness of its beamsplitter $\ell_\text{BS}$,\footnote{The effective thickness accounts for the differential optical path length through the beamsplitter, including effects due to the index of refraction and the non-normal angle of incidence~\cite[Eq. 5]{Vermeulen:2021epa}.} the equivalent DM strain noise is ${h}_\text{DM}(t) = \mathcal{G}_\text{arm} (L_\text{arm}/\ell_\text{BS})  h_\text{GW} (t)$,
where the gain $\mathcal{G}_\text{arm}$ of the arm cavities decreases the overall sensitivity to fluctuations in the beamsplitter thickness~\cite{Grote:2019uvn}. %

In \cref{fig:DM strain spectra} we show the strain noise referred to the DM strain for the Livingston detector during O3. We also show the sensitivity of the GEO600 detector, which has a better sensitivity to DM because of the thicker beamsplitter, higher level of squeezing, and absence of arm cavities. On the same figure, we also show estimated noise performance of the LIGO A+ configuration, as well as the as future gravitational-wave detector, Cosmic Explorer (CE)~\cite{Evans:2021gyd}. We provide a summary of the most important experimental parameters to compare the performance of GW detectors in \cref{tab:gw detector params}. It should be noted that the GEO600 projection assumed a single detector, while the others assumed a cross-correlation of two detectors.
In the case of a single detector, the statistical variance in the noise background limits the minimum detectable value of $\mathfrak{h}_\text{DM}$ to
\begin{equation}
\label{eq:dm_limit_single}
    \mathfrak{h}_\mathrm{DM}\le \frac{\sqrt{\rho \, S(\Omega_\text{DM})}}{(T \tau_\text{DM})^{1/4}},
\end{equation}
where $S(\Omega)$ is the detector noise expressed as an equivalent dark-matter strain spectrum; this limit has the same scalings as the cross-correlation limit \cref{eq:dm_limit_xcorr} and differs only by order-unity factors. In \cref{fig:DM strain} we show the projected upper limits from these experiments. %
Finally, we note that other ground-based gravitational-wave detectors like Virgo~\cite{VIRGO:2014yos} and Kagra~\cite{KAGRA:2020tym} could likely be used in a similar fashion as LIGO, GEO600, and Cosmic Explorer to search for scalar dark matter.

\begin{table}[t]
\caption{\label{tab:gw detector params}Parameters for scalar-field dark matter detection using the beamsplitters of laser interferometric gravitational-wave detectors. The thickness given here is the physical thickness of the beamsplitter, not its effective optical thickness.}
\begin{ruledtabular}
    \begin{tabular}{r l l l}
        & GEO600 & LIGO & CE \\
    \hline
    Beamsplitter thickness [cm] & 8 & 6 & 6 \\
    Arm length [km] & 1.2 & 4 & 40  \\
    Arm gain & 1 & 280 & 280 \\
    \end{tabular}
\end{ruledtabular}
\end{table}

\section{Space based experiments}

Space-based missions such as LISA may also employ solid reference cavities as part of a laser frequency stabilization scheme.
Similar to the case with the LIGO solid cavities, a DM search can be carried out by comparing the cavity-stabilized laser light to the light propagating between the freely falling LISA spacecraft.
We can also project the sensitivity of LISA using the reference cavity stability requirement~\cite{Heinzel:2006un,Valliyakalayil:2021jxd}
\begin{equation}
    \sqrt{S_{\nu\nu}(\Omega)} = \left(\SI{30}{\Hz\big/\sqrt{\Hz}}\right) \times \sqrt{1 + \left(\frac{\SI{2}{\mHz}}{\Omega/2\pi}\right)^4}
\end{equation}
from \SI{0.1}{\mHz} to \SI{1}{\Hz}.
For $T > \tau_\text{DM}$, the limit to DM detection
follows from previous formulae.
However, for millihertz frequencies, $\tau_\text{DM}$ exceeds several years, which is comparable to the expected mission duration.
Thus for these low search frequencies, the sensitivity to $\mathfrak{h}_\text{DM}$ for either a cross-correlation search or single-detector search will scale with integration time like $T^{1/2}$ rather than $T^{1/4}$~\cite{Graham:2015ifn}.
The projected limits are shown in \cref{fig:DM strain}.
This assumes a total search time $T = \SI{4}{yr}$.

\section{Discussion}

While the exact nature of dark matter remains elusive, concerted efforts to exclude all possible interactions are required.
In this work, we provide a method to search for DM interacting with the LIGO detectors\,---\,%
via the length modulation of the solid laser-frequency stabilization cavity\,---\,in addition to previously discussed interactions in the main interferometer's beamsplitter~\cite{Grote:2019uvn}. We find that these methods can be competitive with previous limits in the \(\SI{1e-13}{\eV}/c^2\) mass range even with O3 sensitivity.

In order to run this search, we propose using about one year of data, binning it with longer than usual FFT segments in order to optimize sensitivity in the \SIrange{10}{90}{\Hz} frequency band.
The solid cavity readback channel will need to be tagged for glitches and coherence in order to run this search.
Finally, akin to other cross-correlation and GW searches, the data will be used from time segments that are glitch free and low-noise in both detectors simultaneously.

We also show prospects to obtain more stringent limits by future experiments.
Firstly, we propose lowering of technical noise in the Advanced LIGO frequency locking loop for the next observing runs.
We show the possible DM coupling limits if the LIGO solid cavity could be operated at its thermal noise limit.
We compare this limit with an upgraded cavity with longer length and lower loss mirror coatings.
Finally, we show that the beamsplitter in CE, at its currently expected thickness would provide marginally more stringent limits than the Advanced LIGO beamsplitter, but will not be competitive with limits obtained from thermal-noise limited spacer cavities.
It is notable that if the current LIGO reference cavities can be operated at the thermal noise limit, they will provide the best constraints up to 100 Hz in the absence of lower thermal noise spacer cavities.

\begin{acknowledgments}
    EDH is supported by the MathWorks, Inc., and the LIGO Laboratory.
    NA is supported by NSF grant PHY--1806671 and a CIERA Postdoctoral Fellowship from the Center for Interdisciplinary Exploration and Research in Astrophysics at Northwestern University.
    The authors thank Lee McCuller, Craig Cahillane, Matthew Evans, Lisa Barsotti, Max Isi, Keith Riles, and Masha Baryakhtar for valuable discussions.
    LIGO was constructed by the California Institute of Technology and Massachusetts Institute of Technology with funding from the National Science Foundation, and operates under Cooperative Agreement No. PHY--0757058. Advanced LIGO was built under Grant No. PHY--0823459.
    
\end{acknowledgments}

\bibliography{references}

\appendix
\section{DM signal properties}

\emph{Modulation strength}: 
The expected strength of modulation of fundamental constants has been postulated in Refs.~\cite{Arvanitaki:2014faa,Arvanitaki:2015iga,Geraci:2018fax} as
\begin{align}
    A_{e,m_e} &\sim d_{e,m_e}\frac{\hbar}{m_\text{DM}c^3}\sqrt{8\pi \varrho_\text{DM} G}, \label{eq:Aanalytical}\\
    &= 1.7 \times 10^{-18} \times d_{e,m_e} \times \left(\frac{\SI{100}{Hz}}{\Omega_\text{DM}/2\pi}\right) \label{eq:htodinfrequency}\\
    &= 7\times 10^{-18} \times d_{e,m_e} \times \left( \frac{\SI{1e-13}{\eV}/c^2}{m_\text{DM}}\right),
\end{align}
where, \(\rho_\text{DM}\) is the local dark matter density, \(m_\text{DM}\) is the mass of the dark matter particle, and \(\Omega_\text{DM}=\hbar/m_\text{DM}c^2\) is the dominant frequency of modulation.

\emph{Lineshape}: The DM velocity profile in the rest frame of the galaxy is Maxwell--Bolzmann-distributed with a characteristic velocity $v_\text{DM} =\xi c$, with $\xi \simeq \num{e-3}$ and hence a coherence time $\tau_\text{DM} = 1/\xi^2 \Omega_\text{DM}$.
This velocity dispersion leads to a Doppler-broadened lineshape~\cite{Derevianko:2016vpm}
\begin{multline}
    F(\Omega) = \frac{1}{\sqrt{2\pi}} \frac{\tau_\text{DM}}{\eta} \mathrm{e}^{-\eta^2} \mathrm{e}^{-(\Omega-\Omega_\text{DM}')\tau_\text{DM}} \\
        \times \sinh\left[\eta\sqrt{\eta^2 + 2(\Omega - \Omega_\text{DM}')\tau_\text{DM}}\right],
\end{multline}
where $\hbar\Omega_\text{DM}' = \hbar\Omega_\text{DM} + m_\text{DM} v_\text{gal}^2 /2$ results from the Doppler shift due to the velocity $v_\text{gal}$ of the solar system relative to the galactic rest frame, and $\eta = v_\text{gal} / v_\text{DM}$~\cite{Derevianko:2016vpm}.

\section{Noises}

\emph{Thermal noise:} Given some assumptions\,---\,namely that the coating and substrate have similar mechanical parameters and that the internal mechanical loss angles of the mirrors are similar for bulk and shear modes\,---\,the power spectral density of the Brownian motion in a single mirror is~\cite{Hong:2012jv} 
\begin{equation}
    S_{\nu\nu}^\text{(cBr)}(\Omega) = \frac{8 k_\text{B} T}{\pi \Omega} \times \left(\frac{\nu_0}{L_\text{rc}}\right)^2 \times \frac{(1-2\sigma)(1+\sigma)}{E w^2} \times d \times \phi,
\end{equation}
where the relevant parameters for the fused silica cavity are defined in \cref{tab:refcav params}.

\emph{Shot noise}: So long as the light incident on the cavity is coupled into the cavity with high visibility, the frequency-referred shot noise is~\cite{Chalermsongsak:2014aua}
\begin{equation}
    S_{\nu\nu}^{\text{(shot)}}(\Omega) = \frac{3 \pi \hbar c^2 \nu_0}{64 L_\text{rc}^2 \mathcal{F}_\text{rc}^2 P_\text{rc}}
\end{equation}
for $\Omega \ll c/L_\text{rc} \mathcal{F}_\text{rc}$. Parameter definitions and typical values are provided in \cref{tab:gw detector params}.

\section{LIGO frequency loop architecture}
\label{sec:loop algebra}

\begin{figure}
    \centering
    \includegraphics[width=\columnwidth]{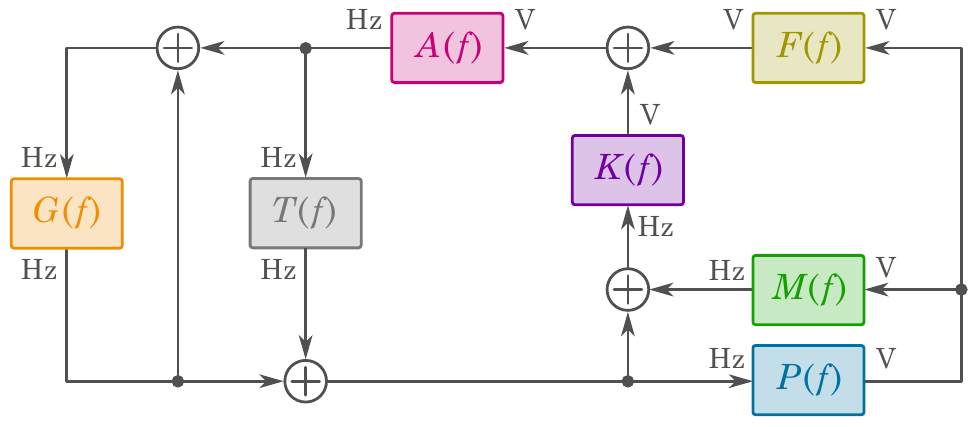}
    \caption{LIGO laser frequency locking loop architecture (compare \cref{fig:carm_diagram}).}
    \label{fig:carm loop}
\end{figure}

\cref{fig:carm loop} shows a loop diagram of LIGO's laser frequency suppression servo system.
In the frequency domain, frequency actuation $a(\Omega)$ applied to the reference cavity error point are found via
\begin{align}
    a &= KA n_\text{mc} + PA(F + MK) n_\text{ifo} \\
      &\hphantom{=} \hspace{3em} + \overline{G}_\text{rc} A K (1 + PF/K + PM) (n_\text{rc} + a) \\
      &=  \frac{KA}{1 - H} n_\text{mc} + \frac{PA(F + MK)}{1 - H} n_\text{ifo} + \overline{H} n_\text{rc}
\end{align}
with $H = \overline{G}_\text{rc} A K (1 + PM + PF/K)$ and $\overline{X} = X/(1-X)$.
Here we have ignored the effect of the tidal servo $T$ in \cref{fig:carm loop}, which is irrelevant for $\Omega/2\pi > \SI{1}{\Hz}$.

The reference cavity gain $G_\text{rc}$, the modecleaner gain $G_\text{mc} = \overline{G}_\text{rc} AK$ and the interferometer gain $G_\text{ifo} = \overline{G}_\text{mc} P (F/K + M)$ are known to satisfy $|G_\text{rc}(\Omega)| \gg |G_\text{mc}(\Omega)| \gg |G_\text{ifo}(\Omega)| \gg 1$ for $\Omega/2\pi < \SI{5}{\kHz}$.
Thus within this frequency band, $\overline{G}_\text{rc} \simeq \overline{G}_\text{mc} \simeq -1$, so $G_\text{mc} \simeq -AK$ and $G_\text{ifo} \simeq -P(F/K+M)$.
Together these imply $H \simeq -G_\text{mc} G_\text{ifo}$, with $|H| \gg 1$ in the relevant frequency band.
Substituting these approximations in 
\begin{equation}
    a \simeq - n_\text{rc} - \frac{n_\text{laser}}{G_\text{rc}} -\frac{n_\text{mc}}{G_\text{ifo}} + n_\text{ifo}.
\end{equation}

\section{Limits}

Comparing \cite[Eq.~4]{Grote:2019uvn} and \cite[Eqs.~(3--4)]{Arvanitaki:2015iga}, we see that $1/\Lambda_e = \sqrt{4\pi} d_{m_e}$ in natural units.
Restoring units, we have $\sqrt{\hbar c^5 / G} / \Lambda_e = \sqrt{4\pi} d_{m_e}$, with $\sqrt{\hbar c^5 / G} = \SI{1.22e19}{\giga\eV}$.
Thus the limit $\Lambda_e > \SI{3e19}{\giga\eV}$ for $m_\text{DM} \sim \SI{1}{\pico\eV}$ from \citet{Vermeulen:2021epa} corresponds to $d_{m_e} < 0.11$.

\end{document}